\documentclass[oneside]{article}
\usepackage[T1]{fontenc}
\usepackage{authblk}
\usepackage{float}
\usepackage{amsmath}
\usepackage{graphicx}
\usepackage{xfrac}
\usepackage[english]{babel}
\usepackage{lmodern}
\usepackage[T1]{fontenc}
\usepackage[latin9]{inputenc}

\usepackage{csquotes}

\usepackage{mathtools}
\usepackage{graphicx}
\usepackage{esint}
\usepackage[unicode=true,
 bookmarks=false,
breaklinks=false,pdfborder={0 0 1},backref=false,colorlinks=false]
 {hyperref}
\usepackage[a4paper, total={210mm,297mm},margin=2.0cm]{geometry}

\usepackage{amssymb}

\usepackage{xcolor}

\usepackage{algorithm}
\usepackage[noend]{algpseudocode}
\usepackage{caption}

\title{A multi-resolution mesoscale approach for microscale hydrodynamics}

\author[1]{Andrea Montessori \thanks{Electronic address: \texttt{a.montessori@iac.cnr.it}; Corresponding author}}
\author[2]{Adriano Tiribocchi}
\author[1]{Marco Lauricella}
\author[2]{Fabio Bonaccorso}
\author[2,1,3]{Sauro Succi}

\affil[1]{Istituto per le Applicazioni del Calcolo CNR, via dei Taurini 19, Rome, Italy}
\affil[2]{Center for Life Nano Science@La Sapienza, Istituto Italiano di Tecnologia, 00161 Roma, Italy}
\affil[3]{Institute for Applied Computational Science, Harvard John A. Paulson School of Engineering And Applied Sciences, Cambridge, MA 02138, United States}

\date{\today}

\begin{document}

\maketitle

\begin{abstract}

We present a new class of multiscale schemes for micro-hydrodynamic problems based on a dual representation of the
fluid observables.

The hybrid model is first tested against the classical flow between two parallel plates and then applied to
a plug flow within a micron-sized striction and a shear flow within a micro-cavity.

Both cases demonstrate the capability of the multiscale approach to reproduce the
correct macroscopic hydrodynamics also in the presence of refined grids (one and two levels), while retaining
the correct thermal fluctuations, embedded in the multiparticle collision method.

This work provides the first step towards a novel class of fully mesoscale hybrid
approaches able to capture the physics of fluids at the micro and nano scales
whenever a continuum representation of the fluid falls short of providing the complete physical
information, due to a lack of resolution and thermal fluctuations.
\end{abstract}


\section{Introduction}

Complex flow simulations across scales
represent a grand-challenge in computational physics and set a ceaseless quest for novel and efficient computational methods.

The main computational challenge is represented by the spatial and time scale separations involved in a plethora of complex systems, from biology to soft matter, which can be addressed by developing efficient hybrid approaches able to bridge these spatiotemporal gaps.

These methods, generally referred to as multiscale approaches, are based on composite computational schemes relying on different levels of description of the system at hand.

Many attempts have been deployed so far in order to tackle this problem, and a number of concurrent hybrid approaches have been proposed to investigate a vast spectrum of physics problems, including soft matter and molecular
fluids,
fluctuating hydrodynamics, as well as both dilute and dense hydrodynamics \cite{delgado2004hybrid,distaso2016,wijesinghe2004three,praprotnik2006adaptive,praprotnik2008multiscale,fedosov2011multiscale}.

In concurrent approaches, most of the physical domain is solved by employing a macroscopic (continuum) description, while only small selected portions are investigated through particle-based methods.

The coupling and communication across the different regions take place within a handshaking region where the hydrodynamic information is exchanged between the two representations.

In this work, we propose a hybrid numerical scheme based on the lattice Boltzmann method (LB) \cite{montessori2019mesoscaleJFM,succi2018lattice,lauricella2018entropic,montessori2015lattice} and the multiparticle collision dynamics (MPCD) \cite{gompper2009multi,kapral2008multiparticle}, capable of predicting the correct macroscopic hydrodynamics also in the presence of multilevel grids (one and two levels), while retaining
the correct thermal fluctuations, embedded by default in the multiparticle collision method.

This work is intended as a very first step towards the development of a novel class of hybrid and fully mesoscale approaches for complex flows across scales, designed in such a way as to capture the physics at the smallest scales, whenever a continuum description alone falls short of providing the correct physical information due to a lack of resolution and thermal fluctuations.

\section{Methods}

\subsection{Lattice Boltzmann method}

The lattice Boltzmann method \cite{benzi1992} (LB) is based on a minimal version of the Bathnagar-Gross-Krook equation, in which  a discrete set of probability distribution functions, flies freely along the links of a cartesian lattice and collide on each node, relaxing towards a Maxwell-Boltzmann equilibrium.
The LB equation reads as follows:

\begin{equation}
	f_i(\vec{x}+\vec{c}_i\Delta t, t + \Delta t)=f_i(\vec{x},t) + \frac{\Delta t}{\tau} [f_i^{eq}(\rho,\vec{u}) - f_i(\vec{x},t)] 
\label{LB}
\end{equation}

where $f_i(\vec{x},t)$ is the discrete distribution function representing 
the probability of finding a particle at position $\vec{x}$ and time $t$, 
streaming along the $i-th$ lattice direction with a (discrete) velocity 
$\vec{c}_i$, $i$ running over the lattice directions (i=0,...,b) \cite{succi2018lattice,montessori2018lattice}. 
The lattice time step $\Delta t$ is usually set to unity as well as the lattice spacing $\Delta x=1$.

The left hand side of the eq.\ref{LB} codes for the streaming process, namely the free-flight of particles along the lattice directions hopping from a site to the neighboring one.

The right hand side implements the relaxation of the set of discrete distributions towards a local thermodynamic equilibrium $f_i^{eq}$, namely a truncated, low Mach number expansion of the Maxwell-Boltzmann distribution \cite{succi2018lattice,kruger2017lattice}.

In this work,a second-order isotropic nine speed two dimensional lattice ($b=8$, $D2Q9$ lattice) has been employed, equipped with a set of second-order expansion of the Maxwell-Boltzmann distribution function, which read as:

\begin{equation}
f_i^{eq}= w_i\rho\left[1 + \frac{\vec{c}_i \cdot \vec{u}}{c_s^2} + \frac{(\vec{c}_i \cdot \vec{u})^2}{2 c_s^4} - \frac{\vec{u} \cdot\vec{u}}{2 c_s^2}\right]
\label{mbeq}
\end{equation} 
In the above equation, $w_i$ are the weights of the discrete equilibrium distribution functions, $c_s^2 = \sum_{i} w_i \vec{c}_{ix}^2=1/3$ the squared lattice speed of sound and $\vec{u}$ the fluid velocity.
The relaxation parameter $\tau$ in equation (\ref{LB}) is linked to the fluid kinematic viscosity via the linear relation $\nu=c_s^2(\tau - \Delta t/2)$ \cite{succi2018lattice}.

The local (time-space) hydrodynamic moments, up to the order supported by the lattice in use are the density ($\rho$), linear momentum ($\rho\vec{u}$) and momentum flux tensor ($\Pi$) \cite{chen2008discrete} and are evaluated by computing the statistical moments of the discrete set of distribution as $\rho= \sum_{i} f_i$, $\rho \vec{u}=\sum_{i} f_i\vec{c}_i $ and $\Pi=\sum_{i} f_i(\vec{c}_i \vec{c}_i-c_s^2 \text{I})$, being $\text{I}$ the identity matrix.

\subsection{Multiparticle collision dynamics with Andersen Thermostat}

In the multiparticle collision dynamics, the fluid is modeled via a large number of pointlike particles of mass $m$, typically of the order of $10^{3}-10^{5}$ in two dimensions \cite{gompper2009multi,ihle2001stochastic}.
During each time step, the system evolves via the subsequent application of a streaming and a collision step.
In the streaming step, the particles positions are updated via a forward Euler step:

\begin{equation}
\vec{r}_k(t + \delta t)=\vec{r}_k(t) + \vec{v}_k(t)\delta t
\end{equation} 
being $\vec{r}_k$ the vector position and $\vec{v}_k$ the velocity of the $k-th$  particle and $\delta t$ the value of the MPCD discrete time step.

As per the collision, several approaches are available \cite{noguchi2007particle,ihle2006consistent,allahyarov2002mesoscopic}.

A possible choice is to perform stochastic rotations of the particle velocities relative to the center-of-mass momentum.
In this case, the multiparticle collision model is usually referred to as  Stochastic Rotation Dynamics (SRD) \cite{malevanets1999mesoscopic}.

In the SRD, the domain is divided into cells of side $a$ \cite{gompper2009multi}.
Multi-particle collisions are then performed within each cell, by rotating the velocity, $\vec{\upsilon}_{k}$, of each $k$-th particle with respect to the velocity of the cell center of mass, $\vec{\upsilon}_{cm}$, of all particles in the cell:

\begin{equation}
\vec{v}_k(t+\delta t)= \vec{v}_{cm}(t) + \mathcal{R}\left[\vec{v}_k(t) - v_{cm}(t)\right]
\label{srdcol}
\end{equation} 
 
In the above equation, $\mathcal{R}$ is the rotation matrix, which rotates the particle velocities of a given cell by an angle $\pm \alpha$ with uniform probability $[\sfrac{1}{2},\sfrac{-1}{2}]$.

The kinematic viscosity of the SRD fluid \cite{kikuchi2003transport} which, in two dimensions, reads as follows:
\begin{equation}
\begin{split}
\nu= \frac{N k_BT\delta t}{a^2}\left[ \frac{N}{(N - 1 + e^{-N})(1 - cos(2\alpha))}\right] + \\
\frac{m (1-cos(\alpha))}{12 \delta t}(N - 1 + e^{-N})
\end{split} 
\end{equation} 
being $N$ is the average number of particles in a collision cell.

A stochastic rotation of the particle velocities is not the only possibility to perform multi-particle collisions. 
In particular, MPCD simulations can be performed directly in the canonical ensemble by employing an Andersen thermostat (MPCD-AT) \cite{gompper2009multi,noguchi2007particle,allahyarov2002mesoscopic}.

In the MPCD-AT, new relative velocities are generated during each computational step in each cell and the collision step writes as \cite{noguchi2007particle}:

\begin{equation}
\vec{v}_k(t+\delta t)= \vec{v}_{cm}(t) + \delta \vec{v}_k^{ran}=\vec{v}_{cm}(t) + \vec{v}_k^{ran} - \frac{1}{N_c}\sum_{j \in cell} \vec{v}_j^{ran}
\label{MPCD-AT}
\end{equation}
where $\vec{v}_k^{ran}$ are random numbers sampled from a Gaussian distribution with variance $\sqrt{k_BT/m}$ and $N_c$ is the actual number of particles in the collision cell.
In the above equation, the sum runs over all particles in a given cell.

In the case of the MPCD with Andersen Thermostat, the kinematic viscosity is given by \cite{gompper2009multi}:

\begin{equation}
\begin{split}
\nu= k_BT\Delta t \left[ \frac{N}{N-1 + e^{-N}} - \frac{1}{2} \right] + \frac{a^2}{12 \Delta t} \left[ \frac{N-1 +e^{-N}}{N}  \right]
\end{split} 
\end{equation}

In this work we employed an MPCD with the Andersen thermostat since, although computationally more expensive than SRD, 
it features shorter relaxation times. This implies smaller number of time steps  required for transport coefficients to reach their asymptotic
values \cite{kapral2008multiparticle}. 
As a consequence the restricition on maximum number of particles per cell ($5-20$), which commonly applies for the SRD, does not apply to the MPCD-AT, where the relaxation times scale as $(\ln N)^{-1}$  \cite{gompper2009multi}.

We wish to point out that the hydrodynamic limit of the multiparticle collision dynamics approach  corresponds to the fluctuating Navier Stokes equation for athermal and weakly-compressible fluids. 
Indeed, it has been previously shown that the MPCD defines a discrete-time dynamics which has been demonstrated to reproduce the correct long-time hydrodynamics \cite{gompper2009multi}.
In addition to the conservation of mass and momentum, MPCD locally conserves energy as well, which enables simulations in the canonical ensemble, i.e. it fully incorporates both thermal  fluctuations and hydrodynamic interactions.
The temperature of the fluid is then constant (apart from fluctuations) within the bulk and, consequently, the present MCPD model cannot describe thermal transport phenomena.
In this sense, the LB approach and MPCD simulation share some features in common.
In particular, in both approaches, the system works at a well defined temperature ($c_s$ in the lattice Boltzmann and $k_B T$ in the MPCD) but  the  MPCD allows the correct incorporation of thermal fluctuations   while in the LB, being a pre-averaged (noise free) version of the lattice gas automata, thermal fluctuations are not included, unless one forces them into the scheme via a stochastic noise \cite{adhikari2005fluctuating}.

\subsection{Coupling Procedure}

In this subsection, we detail the coupling strategy employed to exchange the hydrodynamic information between the lattice Boltzmann and the multiparticle collision dynamics.

\begin{figure}
\begin{center}
\includegraphics[width=10cm]{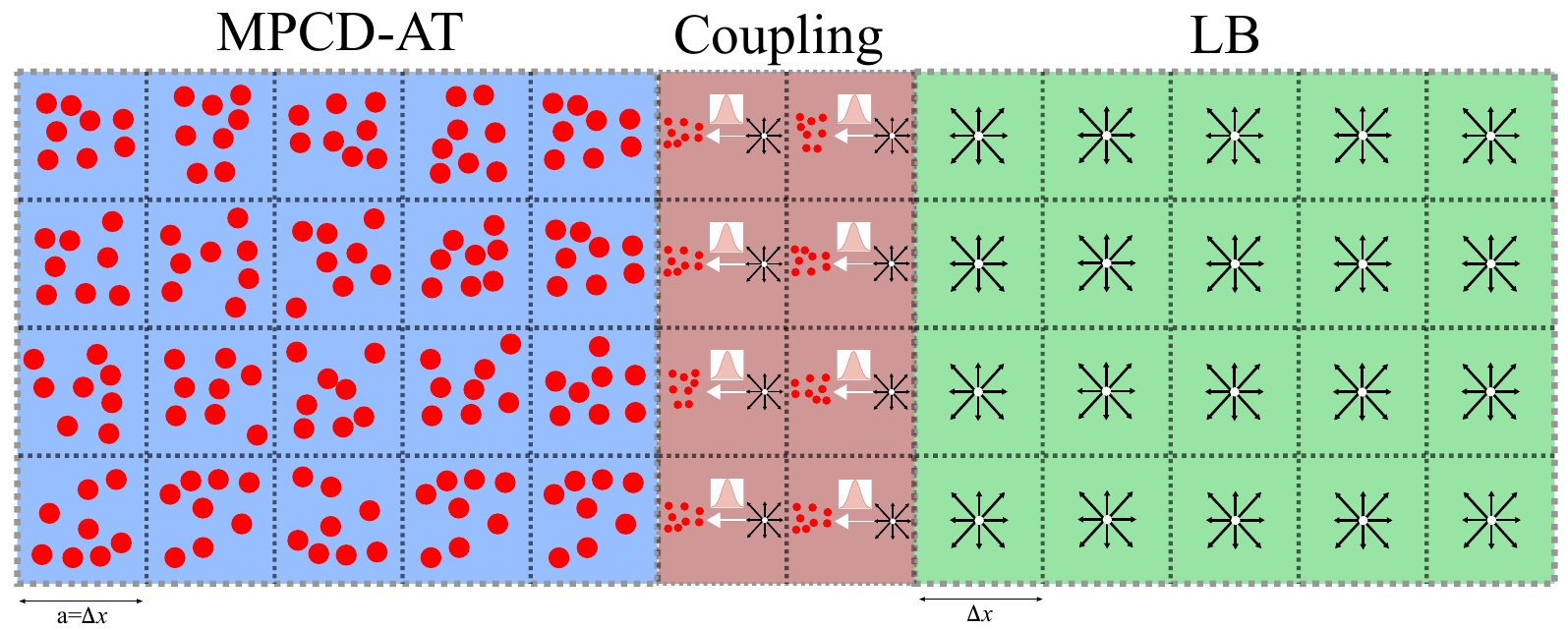}
\end{center}
\caption{Representation of the hybrid domain. On the right region, the pure LB region (green area), and on the left, the high resolution (MPCD, light blue area) region where the streaming collision process of the multiparticle collision dynamics is performed. In the middle, the buffer region (red area) is drawn, where the LB $\to$  MPCD step is performed via the generation of particles velocity from the LB values of the (local) linear momenta.}
\label{fig:1}
\end{figure}

First, we start from the simplest coupling case, namely $\Delta x_{LB}=\Delta x_{MPCD}=a$.
In Figure \ref{fig:1}, we report a sketch of the hybrid simulation domain.
The right region is the pure LB region (green area), while the left one represents the MPCD region (light blue area); 
the red area in the middle is the handshaking region, where the particles velocities are generated from the LB values of the (local) linear momenta.

It is worth noting that, in this work, we employ a one-way coupling, LB to MPCD, between the two approaches \cite{hash1996decoupled}.
The LB runs across the entire domain, the information exchange between the two models is implemented in the coupling region and the MPCD evolves only in selected regions of the domain.

In other words, the information is transferred one-way only, from the LB  to the MPCD domain and, consequently, the LB simulation is unaffected by the noise, which is built-in within the MPCD method.

Moreover, in all the simulations performed, the coupling region covers  up to two lattice nodes, which has proved to be sufficient to obtain smooth transitions of the solutions between the two grids.  

The coupling algorithm proceeds as follows:\\

1) \textbf{LB step}. The LB model runs across the whole domain over a single computational step ( full streaming and collision process ).\\

2) \textbf{Information exchange step}. In the coupling region the particles velocities are re-generated by drawing them  from a normal distribution, generated via the Box-Muller algorithm \cite{box1958note}. 

3) \textbf{MPCD step}. A full streaming and collision step of the MPCD-AT is performed within the MPCD region. 

\begin{figure}
\begin{center}
\includegraphics[width=11cm]{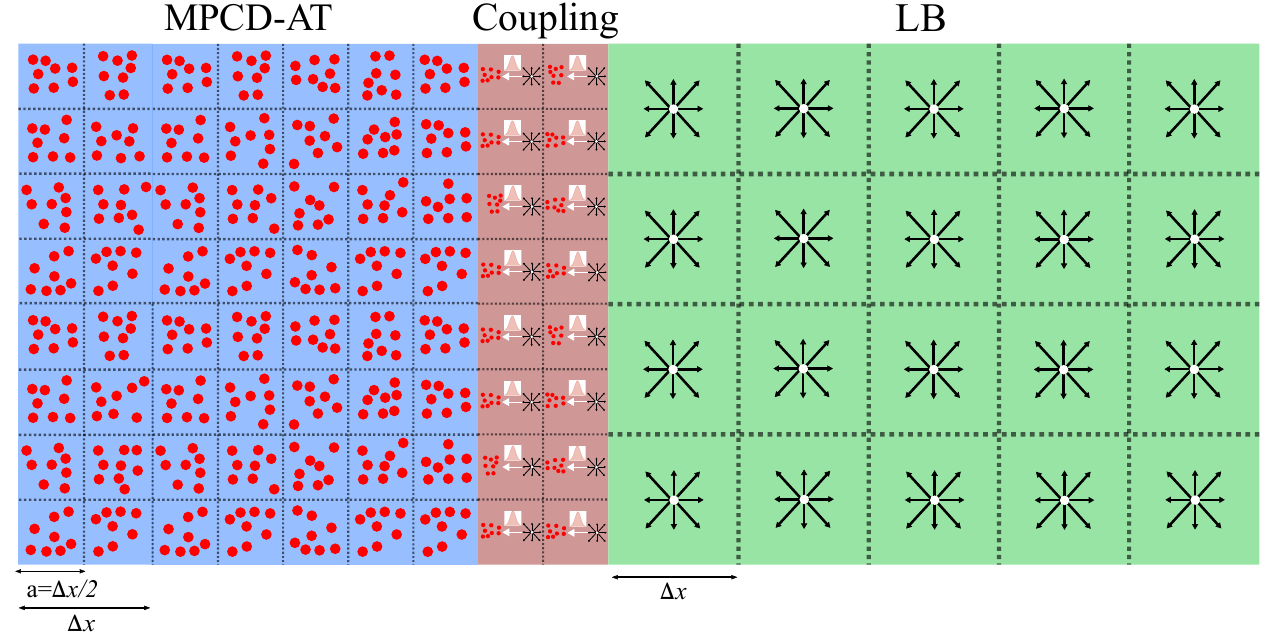}
\end{center}
\caption{Representation of the hybrid multigrid (2 level) domain.}\label{fig:2}
\end{figure}

For the sake of clarity, a pseudo-code is reported in Algorithm \ref{MPCD1}.

\begin{algorithm}
\caption{Coupling procedure : pseudo-code}\label{MPCD1}
\begin{algorithmic}[1]
\Procedure{LB-MPCD coupling}{}
\State \textbf{LB step:}\\
\State \textbf{call} LB collision
\State \textbf{call} LB streaming\\
\State \textbf{Information exchange step:}{}\\
\; \; \;\textbf{for} particles $\in$ coupling region,
	draw $\vec{v}(\vec{x},t)$ from $\frac{1}{\sqrt{2 \pi \frac{k_BT}{m}}}e^{\frac{-(\vec{v} - u)^2}{2\frac{k_BT}{m}}}$\\
\State \textbf{MPCD step}{}
\State \textbf{call} MPCD streaming + boundary conditions
\State \textbf{call} MPCD collision
\EndProcedure
\end{algorithmic}
\end{algorithm}

We then extended the hybrid MPCD-LB approach to run on multigrid domains.  

Specifically,  the LB code runs on the coarse grid, which covers the entire domain, while the MPCD runs on a finer grid with $a=1/2$ or $a=1/4$ (see fig. \ref{fig:2} for a visual sketch of the multigrid hybrid domain with $a=1/2$).

The implementation follows the same procedure outlined in Algorithm \ref{MPCD1} with an additional intermediate interpolation step between the LB and the MPCD steps, where the linear momentum is interpolated from the coarse to the fine grid (see Algorithm \ref{MPCD2}).

\begin{algorithm}
\caption{Coupling procedure : pseudo-code}\label{MPCD2}
\begin{algorithmic}[1]
\Procedure{LB-MPCD two level coupling}{}
\State \textbf{LB step:}\\
\State \textbf{call} LB collision
\State \textbf{call} LB streaming\\
\State \textbf{Bilinear interpolation step:}\\
\For {$i,j$ $\in$ coupling region} $\vec{u}_{coarse} \to \vec{u}_{fine}$
\EndFor\\
\State \textbf{Information exchange step:}\\
\; \; \;\textbf{for} particles in the coupling region,
	draw $\vec{v}(\vec{x},t)$ from $\frac{1}{\sqrt{2 \pi \frac{k_BT}{m}}}e^{\frac{-(\vec{v} - u)^2}{2\frac{k_BT}{m}}}$\\
\State \textbf{MPCD step}{}
\State \textbf{call} MPCD streaming + boundary conditions
\State \textbf{call} MPCD collision
\EndProcedure
\end{algorithmic}
\end{algorithm}

To this purpose, we employed a simple bilinear interpolation.

In the case of the refined simulation the integration time has been chosen so as to avoid sub-cycling in the MPCD region. This 
is possible since the MPCD allows to set both the integration time step and the grid discretization independently. Thus, in the case $a=1/2$ the MPCD, integration time step is set to $\Delta t= 2$, in such a way to guarantee  synchronization between the LB and the MPCD simulation step.

As pointed out above, the coupling procedure is  performed via the generation of  the particle velocities in the overlapping region drawn from a Maxwell-Boltzmann distribution.
This implies that, in the coupling region, the non-equilibrium part of the velocity distribution is set to zero. For the cases investigated in this work, this particular choice turned out to be effective, not compromising the accuracy of the coupling procedure, as detailed in the following.

It is worth noting that this coupling may become inaccurate in the presence of strong velocity gradients, where the non-equilibrium information of the velocity distribution cannot be ignored.

In these cases, different approaches, for example a sampling from an Enskog distribution \cite{garcia1998generation,distaso2016}, would be more appropriate.  

\subsubsection{A note on the computational performances}
It is of interest to compare the computational times associated with the LB and MPCD simulations.

First, we note that a particle update (streaming and collision) takes about $ \sim 0.1 \mu s/particle/step$, comparable to
the time needed to update a lattice unit in a single component LB code, $10 \,MLUPS$ (i.e. Mega Lattice Updates per Second), which is a typical performance of a serial implementation of a single phase, LB code \cite{kruger2017lattice}. 

In the case of the MPCD model, the running time scales roughly linearly with the total number of particles and, in our simulations, it varies between $0.01 s/step\div 0.1 s/step$ with the particle densities ranging between $60 \div 600$ particles per bin.  

It is  clear that an efficient parallelization able to exploit the computational capabilities of the latest supercomputers is mandatory in order to make the coupling approach amenable to large scale simulations.

The performance data reported above refer to a serial implementation of the code, run on a Intel Xeon Platinum 8176 CPU based on the Skylake microarchitecture ($2$ sockets each one with $28$ cores with an installed DRAM of 720 GB ). 
The code was compiled by Intel Fortran Compiler version 18.0.2 with the recent AVX-512 instruction set supported on Skylake architectures.
As per the data structure employed, we opted for a Array of Structure for both the LB and MPCD impolementation (see \cite{shet2013data}). A parallel version (openMP-MPI) of the hybrid approach is currently under development.

\section{Results}

\subsection{ Coupled LB-MPCD-AT: center line coupling and near-wall coupling with grid refinement }

\begin{figure}
\begin{center}
\includegraphics[width=12cm]{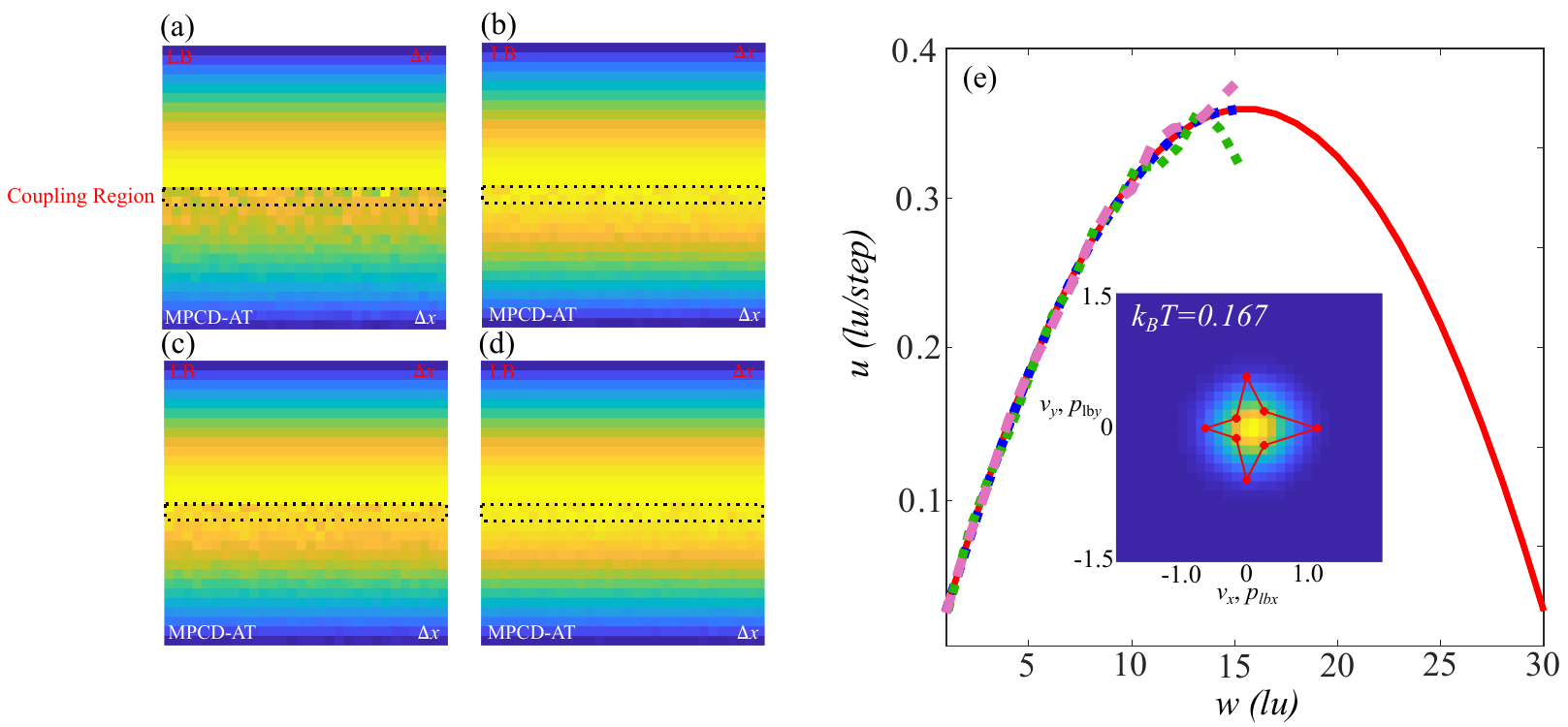}
\end{center}
\caption{(a-d) Hybrid approach, Poiseuille flow velocity field. The coupling region is denoted by dotted region. The cell density of particles was varied between $60 \div 1000$ ((a) to (d)). (e) Plot of the velocity profiles, LB solution (solid line) MPCD (dashed lines) for different values of $N$.  Inset: time-space averaged velocity distribution, comparison between LB(spider plot) and MPCD ($2d$ histogram).  }\label{fig:4}
\end{figure}

We start by testing the coupling procedure for the case of flow between two parallel plates, the main results being shown in figure \ref{fig:4}.

As reported above, the LB runs over a grid covering the entire fluid domain with the MPCD  restricted  to the lower half of the fluid domain,  the depth of the coupling zone extending over two grid nodes,  as denoted by the region included within the dashed line in figure \ref{fig:4}.

Four separate cases have been run for different values of the density of MPCD particles per cell, $N=60 \div 1000$ ((a) to (d)). 

The grids share the same spatial discretization and the same kinematic viscosity, $\nu=0.167\,lu^2/step$ ($lu$ standing for lattice units).

As evidenced in figure \ref{fig:4} (a-d), the LB and MPCD flow fields smoothly connect in the coupling region,  even for the lowest value of particle density per cell ($N=60$). By increasing the cell density, the statistical fluctuations associated to the flow field decrease accordingly, as shown in panel (e) of figure \ref{fig:4} which reports the LB and the MPCD velocity profiles.

In the same plot,a comparison between the continuous (MPCD) and discrete (LB) velocity distribution functions at the steady state is reported (inset of figure \ref{fig:4}(e)). 

The inset displays the time-space averaged velocity distribution.
The MPCD velocity distribution ($2d$ colored field) presents its
typical (shifted) Maxwell-Boltzmann shape while, the LB distributions (spider plot), is
skewed, its skewness increasing for larger values of the mean channel velocity (i.e. as the average Mach number increases).

This departure from the Maxwell Boltzmann behaviour is due to the fact that the set of lattice distributions is not allowed to shift,
as instead occurs in the continuous case, but it occurs via a positive bias in the co-flowing distributions.
For this reason, the larger the non-equilibrium (represented herein
by the large values of the mean channel velocity) the more apparent is the skewness of the discrete set of distributions.

As a further test, we have checked the velocity evolution in two sections of the channel (at $L/2$ and $L/4$, being $L$ the height of the channel) during the transient comparing the results against the analytical solution \cite{morrisjcp1997} and we observed an excellent agreement between  the  numerical and analytical solutions.

We further tested the capability of the hybrid approach to handle multi-resolution grids by performing  simulations of the Poiseuille test with near-wall coupling and grid refinement of the MPCD region .

\begin{figure}
\begin{center}
\includegraphics[width=10cm]{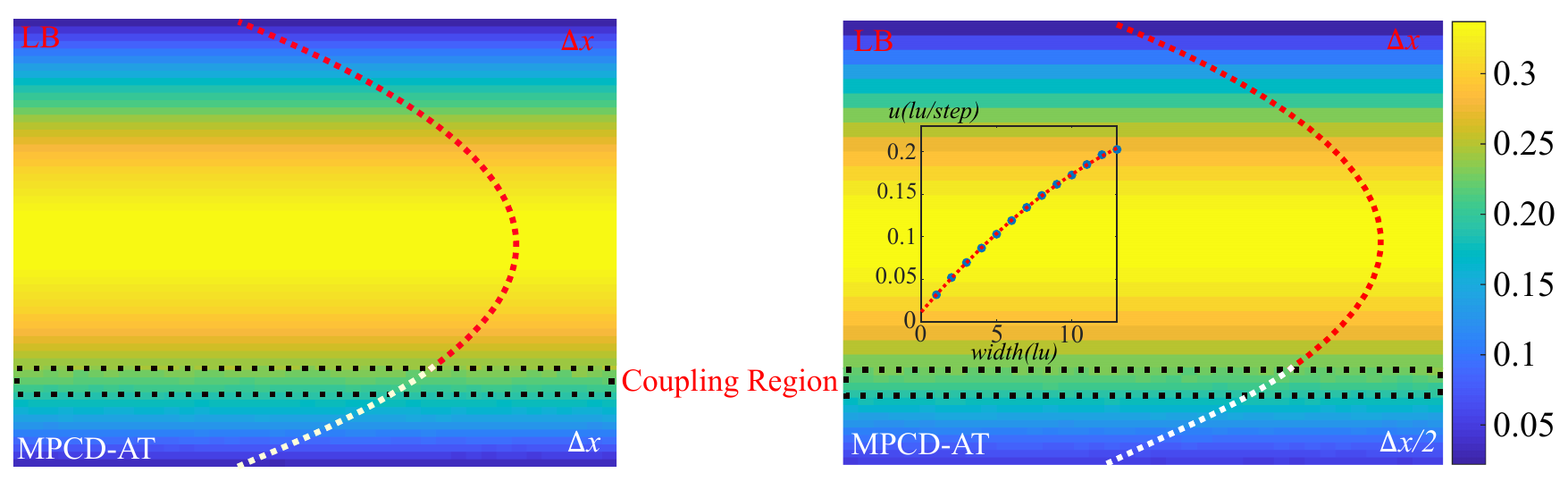}
\end{center}
\caption{(left) Coupled LB-MPCD approach with the same grid discretization (fine grid simulation). (right) Hybrid approach with grid refinement at the wall. Inset : Comparison between the analytical solution of the Poiseuille flow near the wall and the MPCD solution.}\label{fig:5}
\end{figure}

Specifically, we run two separate simulations with the following features:

1)  A reference case where the LB and the MPCD run on a grid with the same discretization (figure \ref{fig:5}(a)). In this case the LB runs on a $40 \times 60$ nodes grid and is coupled to the MPCD near the wall which runs on a $40 \times 14$ bins grid.

2)  In the multi-resolution implementation, the LB runs on a $20 \times 30$ nodes grid ( halved with respect to the previous case), and is coupled near the wall to the MPCD which runs on $39 \times 13$ grid, with a twice finer discretization ($a=1/2$) with respect to the LB grid (figure \ref{fig:5}(b)).

The first case is used as a a benchmark to test the ability of the coupling procedure to correctly reproduce the Poiseuille solution and to handle grid-refined domains.  

In this case, we set $N=1000$ and, as before, the overlapping zone extends to two lattice units.

The main results are reported in figure \ref{fig:5}.

Even in this case, the velocity profiles smoothly reconnect in the overlapping region, which is also quantitatively confirmed by comparing the analytical solution  near the wall against the MPCD  averaged velocity profile, as shown in the inset of fig. \ref{fig:5}, thus proving the effectiveness of the hybrid approach to handle multi-resolution grids.

\subsection{ Plug Flow in a microchannel with a striction}

The multi-resolution algorithm was then employed to simulate the plug flow in a channel with a micrometric striction.

\begin{figure}
\begin{center}
\includegraphics[width=9cm]{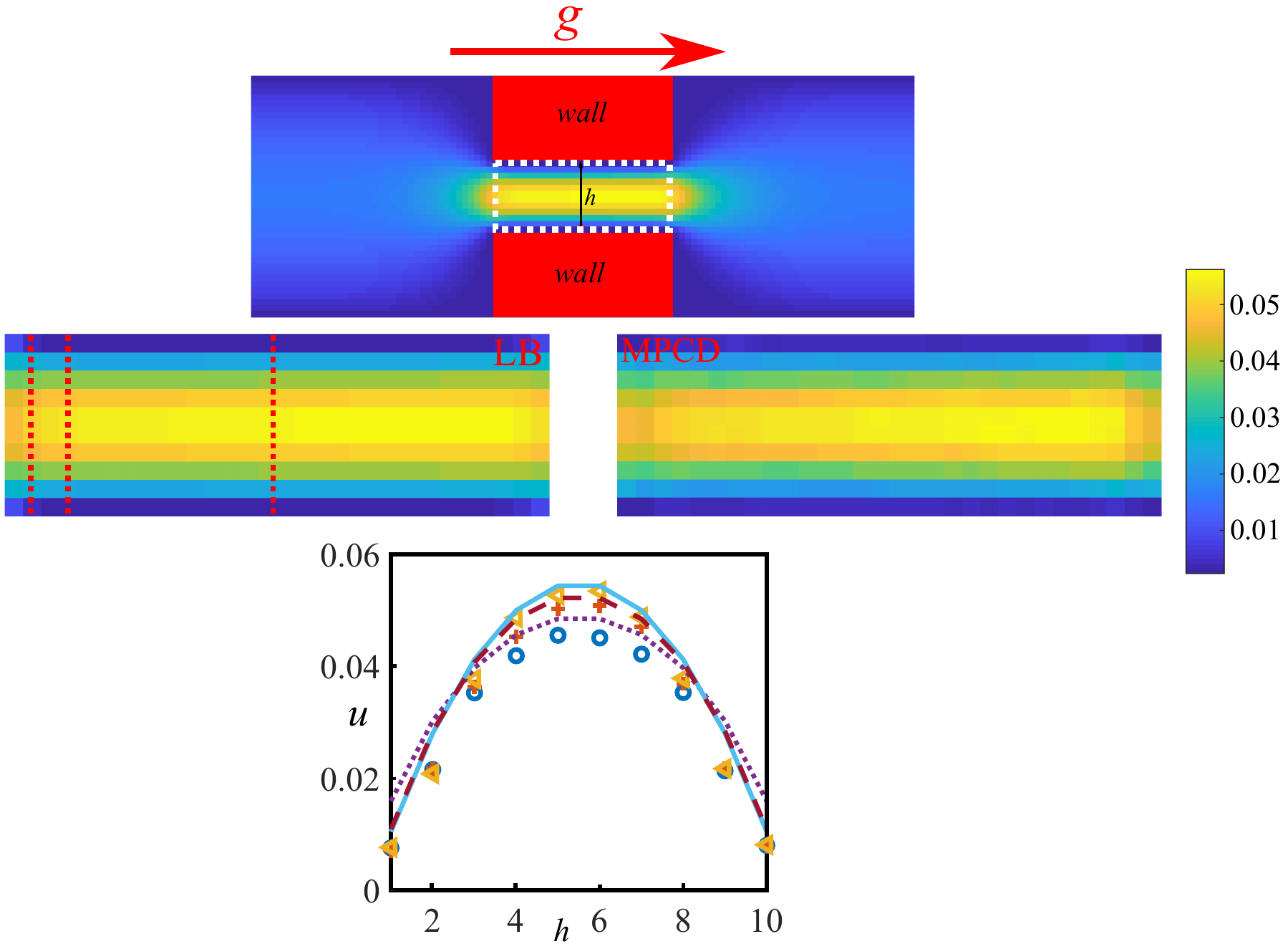}
\end{center}
\caption{ (top panel) Sketch of the flow domain. (center panel) LB (left) and MPCD (right) solutions of the flow field within the microchannel. (bottom panel) comparison between LB and MPCD velocity profiles along the red-dashed vertical lines. Symbols stands for MPCD solutions. Open circles leftmost, crosses central and triangles rightmost section. }\label{fig:6}
\end{figure}

As an example, such geometries,  embedded in more complex microfluidic chips, are widely used for biomicrofluidics applications to perform deformability measurements on red blood cells \cite{pinho2013microfluidic}.
The domain geometry is reported in figure \ref{fig:6}(a). The fluid evolves under the effect of a constant body force ($\vec{g}$) directed from left to right.
In the LB simulation, the domain is periodic along the flow direction, while no-slip boundary conditions are applied both at the upper and lower boundaries and on the striction walls.
In the MPCD sub-domain, denoted by the white dotted rectangular area in figure \ref{fig:6}, no-slip conditions are applied to the particles hitting with the top and bottom walls while, on the right(outlet) and left (inlet) boundaries, particles which would escape from the MPCD domain are periodically re-injected at the opposite side, in order to exactly conserve the mass within the particle domain.

As shown by the colorbar in figure \ref{fig:6}, the maximum flow speed attained within the channel is $\sim 5\cdot 10^{-2}$ (in lattice unit per time step). The channel is discretized with $10$ lattice nodes and the fluid viscosity (in lattice units) is $\nu = 0.167$,
corresponding to a maximum Reynolds number within the channel $Re=\frac{Uh}{\nu}\sim 3$.

By considering a water flow within a $10 \mu m$ height channel, with a maximum speed of $\sim 10 cm/s$, typical values in microfluidic channels,  the Reynolds is  $\mathcal{O}(1)$, thus of the same order as the simulation at hand. 

The LB and MPCD domains are coupled at the inlet and outlet sections of the striction and the overlapping zone extends, as before, over two grid nodes.
Within the coupling regions, the particle velocities are initialized by employing the procedure highlighted in Algorithm \ref{MPCD1}.

Figures \ref{fig:6} (b) and (c) report a visual comparison between the flow fields within the striction obtained with the full LB simulation and with the hybrid approach, while the plot below reports the  velocity profiles taken at three different sections of the microchannel. 
These results confirm that the hybrid approach is able to capture both the dynamical features along the entrance and exit lengths of the striction inlet (see caption in figure \ref{fig:6}) and the bulk dynamics within the channel. 

Moreover, the LB and MPCD velocity profiles show a  fairly good agreement, thus proving the effectiveness of the coupling approach also in the presence of inlet and outlet velocity gradients. 

The same simulations were then performed by discretizing the micrometric channel in the MPCD simulation on finer grids. To this purpose, two sets of simulations have been performed by running the MPCD on a two-fold ($a=1/2$) and four-fold ($a=1/4$) grid resolution.
\begin{figure}
\begin{center}
\includegraphics[width=10cm]{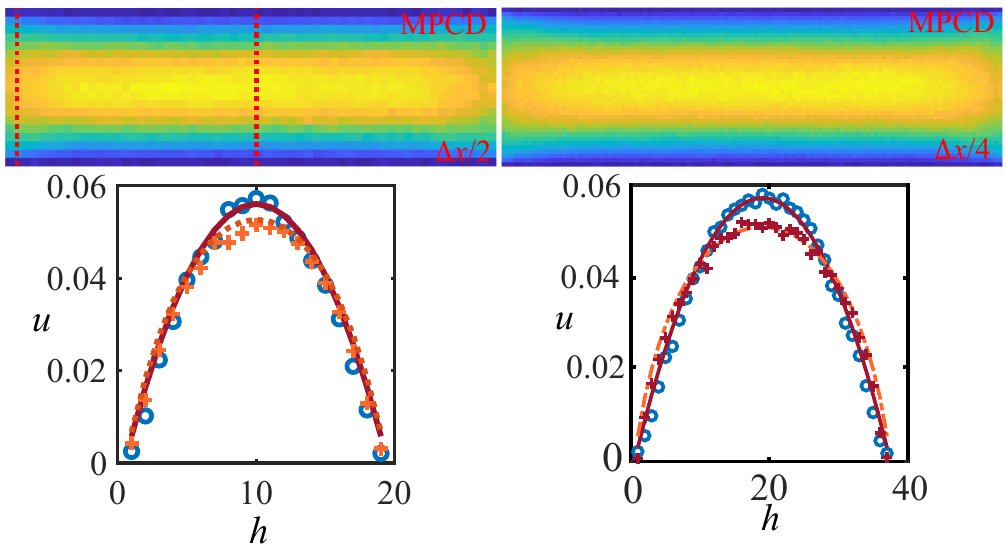}
\end{center}
\caption{ (left and right panels) MPCD velocity fields on grids with increased resolutions with the plot of the velocity along vertical sections denoted by the dashed lines. Symbols denotes MPCD results. }\label{fig:7}
\end{figure}
The main results are reported in figure \ref{fig:7}, showing the flow fields ((a) and (b)) and the velocity profiles along the red-dashed vertical sections of the striction.
In order to compare the MPCD with the LB results, two full resolution LB simulations have been run as a reference.

These results further demonstrate the ability of the multi-resolution approach to correctly reproduce the physics of fluid within the micron-sized channel.

Moreover, as stated in the introduction, the hybrid mesoscale approach not only allows to locally refine the grid to obtain more accurate solutions of the hydrodynamic fields, but also naturally preserves the fluctuating statistics, typical of confined fluids the micro- and nano-scales, which is absent in the non-fluctuating LB models \cite{adhikari2005fluctuating,dunweg2007statistical}. 

\subsection{Shear flow in a channel with a micro-cavity}

As a final case, we present  a shear flow in a periodic channel with a microcavity placed at the bottom. This kind of geometry is typical of micropatterned surfaces \cite{suzuki2015flow} and rough walls in microdevices. 

Similar systems are widely employed in biological and medical applications for selective cell sorting and entrapment, enabling for tunable size-based cell capture in microvortices \cite{khojah2017size}.

\begin{figure}
\begin{center}
\includegraphics[width=5cm]{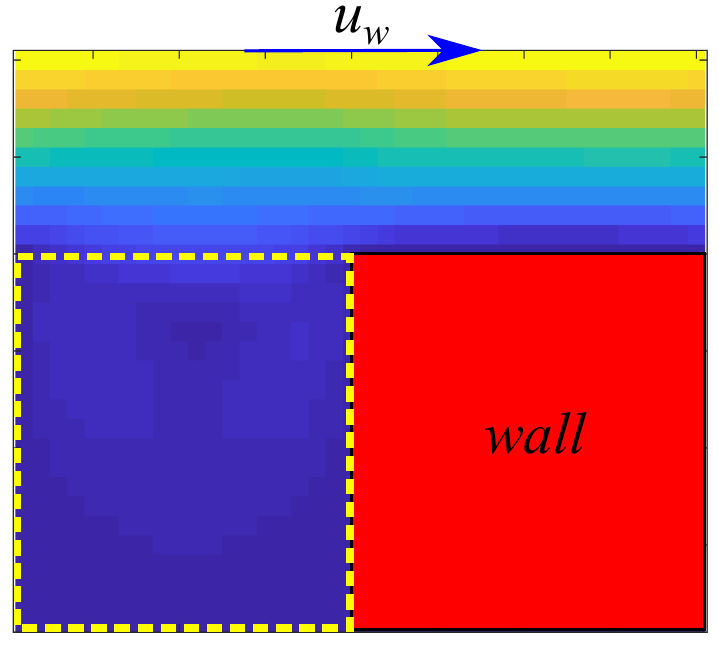}
\end{center}
\caption{ Sketch of the channel with the micro-cavity.}\label{fig:8}
\end{figure}

A sketch of the domain is reported in figure \ref{fig:8}, in which  a shear is applied to the fluid by imposing a momemtum exchange boundary condition to the top boundary \cite{bouzidi2001momentum}. Periodic boundary conditions are applied along the flow directions, while no-slip condition is implemented on the bottom boundary and on the cavity walls.

The main  simulation and geometrical  parameters are the following:

i)  the LB domain is discretized with $40 \times 30$ grid nodes and the square cavity has a side $L=20$. The kinematic viscosity is set to $\nu=0.167$ and the shear velocity imposed at the uppermost boundary is $U=0.4$ in lattice units. 

ii) The cell density in the MPCD simulations is set to $N=1000$. 

iii) As per the boundary conditions employed in the MPCD, the bottom and side walls are no slip, while the uppermost boundary is treated as a free slip wall as the velocity is fixed during the information exchange between the LB and MPCD grids.

It is again of interest to compute the cavity Reynolds number $Re=\frac{UL}{\nu} \sim 6$, typical of micro-cavities under shear \cite{suzuki2015flow,khojah2017size}.
The MPCD run in the region inside the dashed rectangular perimeter (see figure \ref{fig:8})and the coupling region coincides with the separation line between the microcavity and the main channel.
\begin{figure}
\begin{center}
\includegraphics[width=7cm]{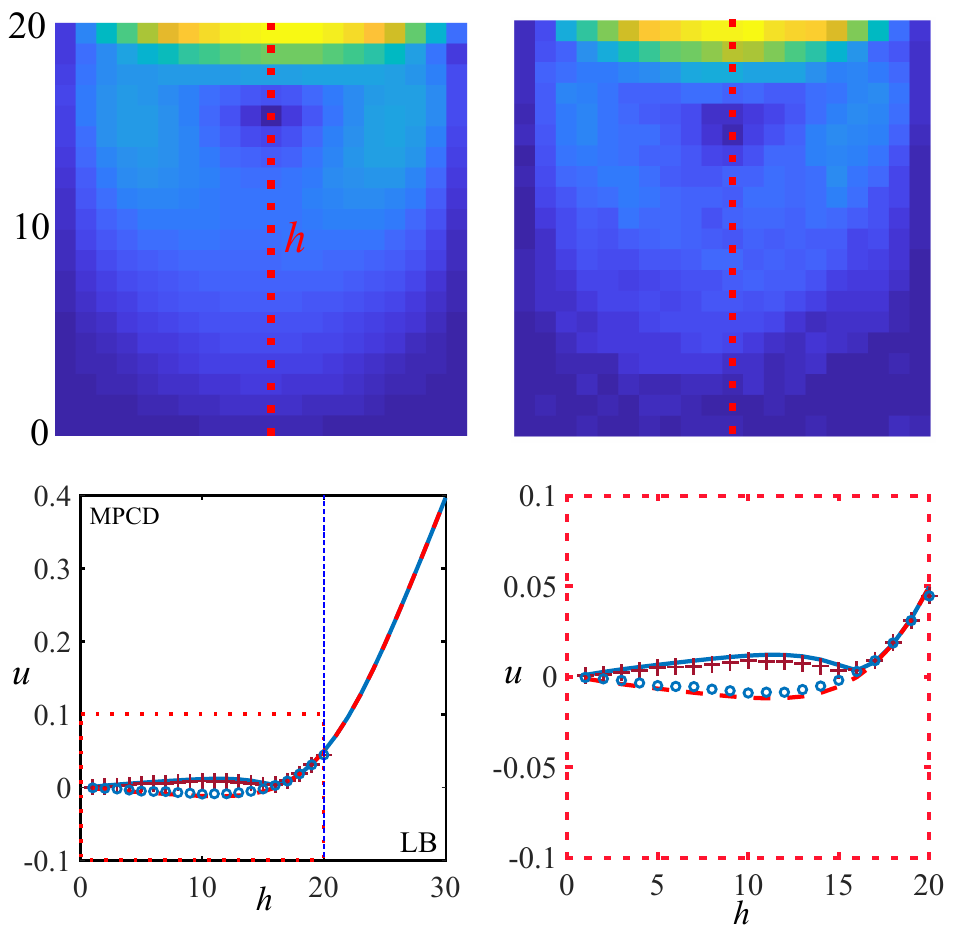}
\end{center}
\caption{ Comparison between the flow fields inside the micro-cavity computed with the LB(left) and the MPCD(right) solver. The plot reports a comparison of the horizontal component of the velocity (open circles and dashed line) and of its magnitude (crosses and solid line),along the vertical axes of the cavity, between the LB and MPCD solutions. Symbols stand for the MPCD solutions.}\label{fig:9}
\end{figure}
As a first case, we run the coupled approach by using the same grid discretization both in the LB and in the MPCD regions.
The first set of results is shown in figure \ref{fig:9}. 
As shown in panel (a) and (b), the main vortex inside the microcavity is accurately captured both in the LB and in the MPCD simulation. 
Further, the vortex positions and sizes in both the representaion also show a good agreement. 

\begin{figure}
\begin{center}
\includegraphics[width=7cm]{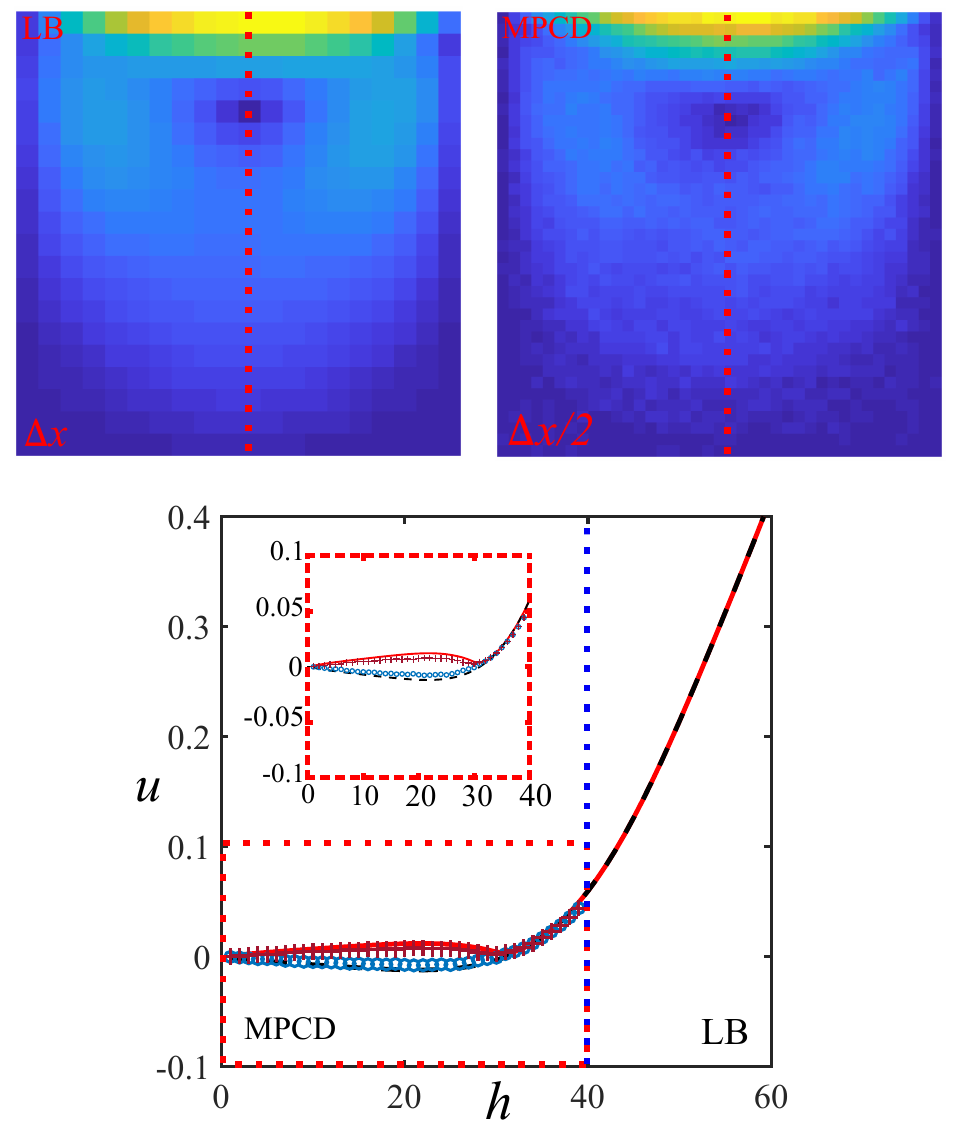}
\end{center}
\caption{  Comparison between the flow fields inside the micro-cavity computed with the LB(left) and the refined MPCD(right). The plot reports a comparison of the horizontal component of the velocity (open circles and dashed line) and its magnitude (crosses and solid line),along the vertical axes of the cavity, between the LB and MPCD solutions. Symbols stand for the MPCD solutions. }\label{fig:10}
\end{figure}

\begin{figure}
\begin{center}
\includegraphics[width=6cm]{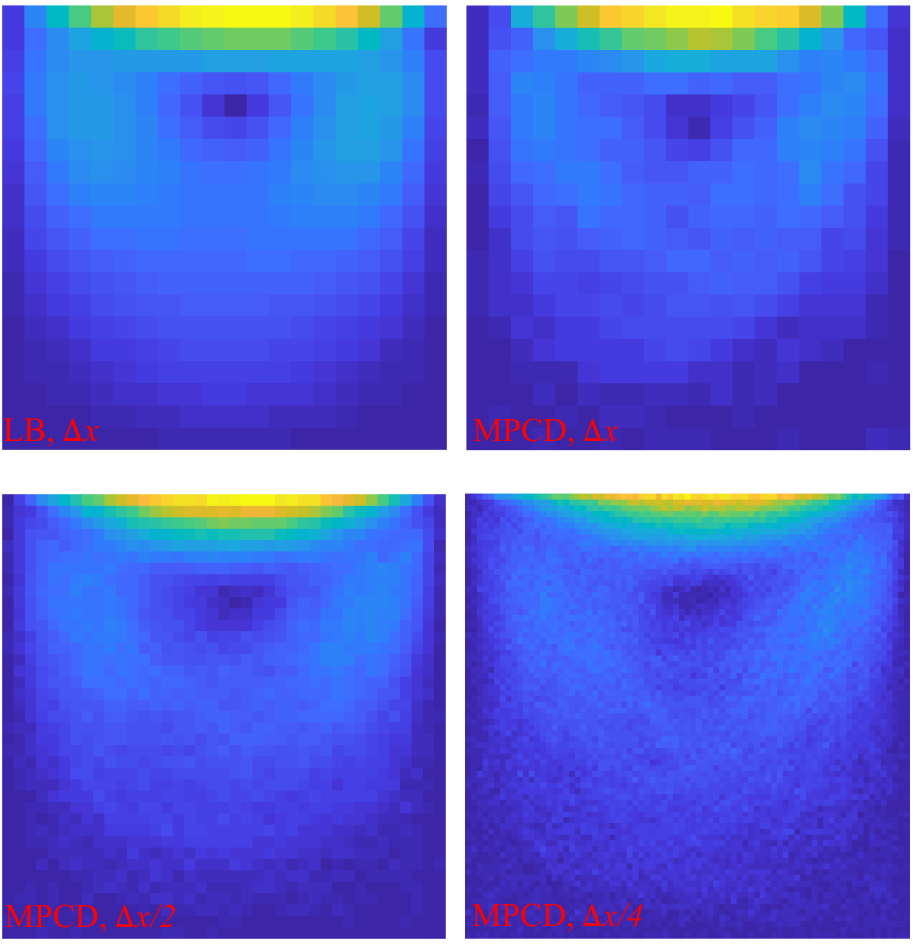}
\end{center}
\caption{Flow fields inside the micro-cavity. Comparison between LB and MPCD with grid refinement.}\label{fig:11}
\end{figure}

We then plotted the horizontal component and the magnitude of the velocity ($\sqrt{u^2 + v^2}$) along the height ($h$) of the cavity (panels (c) and (d)).
The agreement is evident  and confirmes the smooth reconnection between the  MPCD and LB solutions within the coupling region.
Furthermore,the resolution in the MPCD region has been increased by a factor two and four($a=1/2$ $a=1/4$ ). The main results are reported in figures \ref{fig:10} and \ref{fig:11}, which report, respectively, a comparison between the LB and MPCD solutions and the flow fields within the cavity for different values of the grid size.
Here too, the agreement between the LB and the MPCD solution is remarkable, as is also confirmed by the subplot in panel (c), showing the LB and MPCD velocity profiles, taken  along the height of the cavity, in close match with each other.
Also in this case, the multi-resolution approach proves capable of providing the correct macroscopic hydrodynamic information  with  thermal fluctuations directly incorporated in the MPCD solver. 

\section{Conclusions}

In this work a concurrent lattice Boltzmann-multiparticle collision dynamics multiscale/multigrid approach has been developed and validated against the classical laminar flow through parallel plates and employed to simulate fluid flows in confined geometries at the micro-scale.

In particular, for the  Poiseuille flow across two parallel plates, several cases have been investigated, namely 

1) hybrid LB-MPCD with the coupling region positioned at the center of the channel

2) Coupled LB-MPCD  with one level near-wall grid refinement 

Furthermore, the multi-resolution approach has been used to simulate a plug flow in a microchannel with a micrometric striction and a shear flow in a channel with a micro-cavity placed on the bottom.

The results presented are aimed at demonstrating the effectiveness of the hybrid approach to simulate fluctuating Navier-Stokes fluid dynamics even in the presence of  complex geometries and velocity gradients near the coupling region.

As stated in the introduction this work represents a first step towards a fully coupled (MPCD-LB) algorithm,
which is currently under development.
Nevertheless, a one-way coupling, like the one proposed  in this work, can be employed in a number of applications where the effect of thermal fluctuations needs to be taken into account only in small portions of the fluid domain. In this respect, an example is represented by the flow in the microcavity. Indeed, within the microcavity, both hydrodynamic interactions and thermal fluctuations play an important role while outside, in the sheared channel, fluctuations are not essential and a lattice Boltzmann approach can safely be applied.

Further developments of the hybrid approach, currently ongoing, will aim at the simulation of complex fluid phenomena in confined systems, also in the presence of hydrodynamic interactions with nanoparticles, and within thin films at the fluid-fluid interface, wherein increased resolution and thermal fluctuations  are both needed to capture the relevant microscale physics.

\section*{Data Availability}

The code is available upon request.

\section*{Acknowledgements}

A. M., M. L., A. T. and S. S. acknowledge funding from the European Research Council under the European
196 Union's Horizon 2020 Framework Programme (No. FP/2014-2020) ERC Grant Agreement No.739964 (COPMAT).










\end{document}